\documentclass[aps,prl,superscriptaddress,twocolumn,epsf,showpacs]{revtex4}
\usepackage{graphicx}
\begin{document}

\title{Atomistic mechanisms and diameter selection during nanorod growth}
\author{Da-Jun Shu}
\email{djshu@nju.edu.cn}
\affiliation{National Laboratory of Solid State Microstructures and\\
Department of Physics, Nanjing University, Nanjing 210093, China}
\author{Xiang Xiong}
\affiliation{National Laboratory of Solid State Microstructures and\\
Department of Physics, Nanjing University, Nanjing 210093, China}
\author{Zhao-Wu Wang}
\affiliation{National Laboratory of Solid State Microstructures and\\
Department of Physics, Nanjing University, Nanjing 210093, China}
\author{Zhenyu Zhang}
\affiliation{Department of Physics and Astronomy, The University of Tennessee, Knoxville,
Tennessee 37996, USA}
\affiliation{Materials Science and Technology Division, Oak Ridge National Laboratory,
Oak Ridge, Tennessee 37831, USA}
\author{Mu Wang}
\email{muwang@nju.edu.cn}
\affiliation{National Laboratory of Solid State Microstructures and\\
Department of Physics, Nanjing University, Nanjing 210093, China}
\author{Nai-Ben Ming }
\affiliation{National Laboratory of Solid State Microstructures and\\
Department of Physics, Nanjing University, Nanjing 210093, China}

\begin{abstract}
We study in this paper the atomic mechanisms of nanorod growth and propose the way of diameter selection of nanorod. A characteristic radius is demonstrated to be crucial in nanorod growth, which increases proportional to one fifth  power of the ratio of  the interlayer hopping rate of adatoms across the monolayer steps to the deposition rate.  When the radius of the initial island is larger than this characteristic radius, the growth morphology evolves from a taper-like structure to a nanorod with radius equal to the characteristic radius after some transient layers. Otherwise the nanorod morphology can be maintained during the growth, with stable radius being limited by both the radius of the initial island and the three-dimensional Ehrlich-Schwoebel barrier. Therefore different growth modes and  diameter of nanorod can be selected by tuning the characteristic radius. The theoretical predictions are in good agreement with experimental observations of ZnO growth.
\end{abstract}

\pacs{61.46.-w, 68.65.Ac, 68.55.-a}

\maketitle
\section{I. Introduction}
One-dimensional nanostructures have attracted much attention since they provide potential applications for nanoelectronics and nanophotonics \cite{1Dnano,Yang01,Nobis,Vugt,zhchen}.
 For example, zinc oxide nanorods with hexagonal cross-section can be applied as whispering gallery resonators, in which the coupling between the resonant modes and free excitons depends sensitively on the cross-sectional radius \cite{Nobis,Vugt,zhchen}. Growth of nanorods have been extensively reported thus far, yet there are  few studies concentrating on the underlying atomistic mechanisms of growth, especially on the understanding and controlling the cross-sectional radius of nanorods from atomic point of view \cite{wang04,rzf02,Kong,variT,variP,WangRC,Umar06,PLD}.

It is known that surface kinetics plays an important role in determining the morphology and size of nanostructures \cite{Zhang97,Tersoff94,Krug,Villain00}. The deposited adatoms can either diffuse within the topmost layer and aggregate to form a new layer nucleus, or hop downward across the step edges and contribute to the lateral growth of the topmost layer.
Under a certain deposition condition, the kinetics-controlled competition between the growth in the normal direction to the substrate and the lateral growth is expected to determine the growth modes and morphologies \cite{Tersoff94,Krug}.

The   surface kinetics can be described by intralayer and interlayer hopping rates of adatoms, $\nu =\nu _{0}\exp (-E_{d}/kT)$ and $\nu ^{\prime }=\nu'_{0}\exp (-E_{s}/kT)$, respectively, where the prefactors $\nu_0$ and $\nu'_0$ are the attempt rates which are approximately of the same value; $k$ is the Boltzmann's constant and $T$ the temperature.  The interlayer diffusion barrier ($E_{s}$) is normally larger than the intralayer one  ($E_{d}$).  The difference of these two values is denoted as $E_{es}$, which is  known as Ehrlich-Schwoebel barrier (ESB) \cite{ES1,ES2}. The ratio of $\nu/\nu'$ increases with ESB as $\exp(E_{es}/kT)$. The ESB is reported to increase with the step height and saturate in several atomic layers, with a value usually referred  as three-dimensional (3D) ESB. The  conventional ESB is applicable for a monolayer step and is hereafter termed as two-dimensional (2D) ESB \cite{LiuH,Zhang02}.

 When ESB is small enough to allow sufficient inter-layer diffusion, layer-by-layer growth occurs, whereas larger ESB leads to multilayer growth \cite{Tersoff94}. In the latter case, islands initiate from each nucleus, which approach nanorods if the cross-sections remain the same along the longitudinal direction. The key point to select and control the nanorod growth is to understand how the radius of the cross-section varies when atomic layers add up under specific growth conditions. Since there is a large difference between the 2D- and 3D- ESBs, it is also crucial to identify the specific roles of 2D- and 3D-ESBs in the nanorod growth.

In this paper, we study the influence of deposition rate and ESB on the growth process of nanorods. A characteristic radius has been identified, which increases proportional to one fifth  power of the ratio of  the 2D-ESB limited hopping rate of adatoms to the deposition rate. Both the growth modes and the nanorod diameter can be selected by tuning this characteristic radius.
 We demonstrate that when the radius of the initial island is larger than the characteristic radius, the growth morphology evolves from a taper-like structure to a nanorod with radius equal to the characteristic radius. However, if  the characteristic radius becomes larger than the radius of the initial island, by increasing the 2D-ESB limited hopping rate or decreasing  the deposition rate, nanorod morphology can be maintained during the growth, with a stable radius being limited by both the radius of the initial island and 3D-ESB. The theoretical predictions of the characteristic radius  are  demonstrated  with experimental observations of ZnO growth, and good consistency  has been found.

\section{II. Nucleation  on top of an island}

Let us consider a nanostructure with thickness of $n$ atomic layers.  For simplicity, the cross section is taken as circular. The radius of the $i$-th layer is  denoted as  $R_i$, measured in terms of the surface cell parameter $a_0$. The dimensionless area  and perimeter are  $A_i=\pi R_i^2$ and $L_i=2\pi R_i$, respectively.

Assuming the growth units are deposited  in the normal direction of the surface,  at rate $F$  per surface cell of area $a_0^2$. The number of adatom $\eta$ per surface cell is determined by the diffusion euqation,  $d\eta /dt = \nu \nabla^2 \eta +F $. By solving the diffusion equation, the distribution of the number density of adatom can be obtained,
 \begin{equation}
\eta = \frac{F}{4\nu}(R_n^2-r^2)+\eta_e, \label{eq:eta}
 \end{equation}
  where $\eta_e$ is the dimensionless number density of adatom at the edge of $A_n$. Before nucleation occurs on top of $A_n$, a deposited adatom on $A_{n}$ has no other choice but to hop across the  step edges after a survival time of $\tau$. At steady state, the number of atoms leaving the surface  per unit time,  $L_n\eta_e\nu'$, is balanced by that of the atoms deposited on the surface per unit time, $FA_{n}$. It gives the number density of adatom on the boundary,
   \begin{equation}
\eta_e= (FR_n)/(2\nu'). \label{eq:eta-e}
 \end{equation}
   The total number of adatom on $A_n$ can be obtained by integrating,
      \begin{equation}
N= \int_0^{R_n} \eta 2\pi r  dr=\frac{F\pi R_n^3}{2\nu'}(\frac{R_n}{4\nu/\nu'}+1).
 \end{equation}
    The average survival time of an adatom is thus $\tau = N\Delta t$, where $\Delta t=1/(FA_{n})$  is the time interval between subsequent deposition events on $A_{n}$. In addition to $\Delta t$ and $\tau$, another concerned  time scale  is the traversal time for an atom to visit all the sites of $A_{n}$, $\tau _{tr}=A_n/\nu $.

As   $A_n$ grows, the probability of nucleation on $A_n$ increases.  Once a new nucleus forms on $A_n$, the number of atomic layers  $n$ increases by one,  which leads to the growth in the  normal direction  to the substrate. For the simplest case that a dimer is the smallest stable island, the nucleation rate on $A_n$ can be given by $\Omega =p_{1}p_{2}/\Delta t$, where $p_1=1-\exp (-\tau /\Delta t)$ is the probability that an atom is deposited during the presence of another atom on the surface, and $p_2=1-\exp (-\tau /\tau _{tr})$ is the encounter probability \cite{Rottler}.

It has been reported that with slow deposition  the total number of adatoms is usually much less than unity, $i.e.$  $N\ll 1$, which means that $\tau \ll \Delta t$ \cite{Krug,Rottler}. Furthermore, in typical island growth with large ESB,  $\nu/\nu'$ is much larger than dimensionless $R_n$, so the first term in Eq.~(\ref{eq:eta}) is much smaller than the second term, and the number density of adatom on $A_n$ is approximately uniform, $\eta \simeq \eta_e$.  The total number of adatom becomes $N= \frac{F\pi R_n^3}{2\nu'}$, which gives
      \begin{equation}
\tau=N\Delta t = \frac{R_n}{2\nu'}.
 \end{equation}
It indicates that $\tau \gg \tau_{tr}$ when $\nu/\nu' \gg R_n$,  and the encounter probability of two adatoms simultaneously present on the island $p_2$ is approximately  unity.  The nucleation rate therefore can be approximated as
  \begin{equation}
\Omega = \frac{F^{2}A_{n}^{3}}{L_n\nu ^{\prime }} = \frac{F^2 \pi ^2 R_n^5 }{2\nu ^{\prime }}. \label{eq:omega}
    \end{equation}
Depending on the height of the steps across which the adatoms hop to the lower layer, 2D-ESB or 3D-ESB plays the role to influence the interlayer atomic diffusion, respectively. Correspondingly subscripts  $2D$ and $3D$ will be added  to  $\nu'$ or $\Omega$ in the context to show such a difference.


\section{III. Characteristic Radius}
  For a buried layer, $i.e.$ the atomic layer above which a second layer has formed, the condition $\tau \gg \tau_{tr}$ means that an adatom can always be trapped by the ascending steps before getting chance to hop to the lower layer.  The adatoms on $A_n$  before second-layer nucleation, however, has no other choice but to hop to $A_{n-1}$. The lateral growth of the topmost layer $A_n$ is thus contributed by the atoms deposited on area $A_{n-1}$,
\begin{equation}
dA_{n}/dt=FA_{n-1}= F\pi R_{n-1}^2. \label{eq:dA}
\end{equation}
We assume for the moment that $A_{n-1}$ is large enough so that $A_n$ is always smaller than $A_{n-1}$.

  The probability $f$ that a second layer has nucleated on $A_n$, according to $df/dt=\Omega(1-f)$, increases with $A_n$ as the following,  \begin{equation}
  f=1-\exp (-I), \label{eq:f}
     \end{equation}
 where \begin{equation}
  I=\int_{0}^{A_{n}}\Omega_{2D} \frac{dt}{dA}dA \label{eq:I0}
     \end{equation}
  can be regarded as the average number of nucleus on $A_{n}$. Substituting Eqs.\ (\ref{eq:omega}) and (\ref{eq:dA}) into Eq.(\ref{eq:I0}),
 \begin{equation}
 I=\frac{F}{\nu_{2D}^{\prime }}\frac{\pi ^{2}R_{n}^{7}}{7R_{n-1}^{2}}=%
\frac{R_{n}^{7}}{R_{c}^{5}R_{n-1}^{2}}, \label{eq:I}
 \end{equation}
where $R_c$ is defined as
\begin{equation}
R_{c}=\left( \frac{7\nu_{2D}^{\prime }}{F\pi ^{2}}\right) ^{1/5}. \label{eq:L2d}
\end{equation}
 $\nu_{2D}^{\prime }$ in Eq.\ (\ref{eq:L2d}) denotes the hopping rate of adatom from  $A_n$ to $A_{n-1}$, with subscript ${2D}$ emphasizing that $R_{c}$ is determined by the 2D-ESB across the monolayer step edges. Since the probability $f$ goes rapidly from nearly zero  to nearly unity when $I=1$,
the condition that $I=1$ can be used as a criterion for the formation of the $(n+1)$-th layer \cite{Tersoff94}.
According to Eq.~(\ref{eq:I}), the second layer nucleus has formed before $R_{n}$ reaches $R_{n-1}$ if $R_{n-1}>R_{c}$. If $R_{n-1}<R_{c}$, however, the probability that second-nucleus forms on top of $A_n$ is still nearly zero even though $R_n$ reaches $R_{n-1}$.

As $R_{c}$ is determined just by the 2D-ESB and the growth conditions, such as the deposition rate and the growth temperature,  it can be regarded as a characteristic radius of the growth system.
In heterogeneous growth, it is known that the effect of the foreign substrate on surface kinetics properties depends strongly on the thickness of the grown layers. The interlayer hopping rate $\nu_{2D}'$ is therefore variable, especially in the first two layers. Accordingly, we denote hereafter the characteristic radius of the first layer and the other layers as $R_{c0}$ and $R_c$, respectively, in order to distinguish their difference.

It is known that $R_{c0}$ is critical in determination of the growth mode, such as  layer-by-layer growth or island growth in the beginning of heterogenous growth. Here we propose  that, once the island growth sets in, it is the characteristic radius $R_{c}$ that plays a key role during the development of a separate island in selecting growth mechanisms and the lateral size of the island.

\section{IV. Two Growth Scenarios}
In heterogenous growth, the average radius of the foreign substrate occupied by each nucleus is denoted as $R_{0}$. If $R_0$ is larger than  $R_{c0}$, a second-layer nucleus forms when the radius of the first layer approaches $R_{1}=R_{c0}\left( {R_{0}}/{R_{c0}}\right) ^{2/7}<R_0$, which means that the second layer nucleus forms before the first layers coalescence and thus  island growth sets in.  We define $R_1$ as the radius of the initial island from which the island growth  starts.

\begin{figure}[t]
\includegraphics[width=8.5cm]{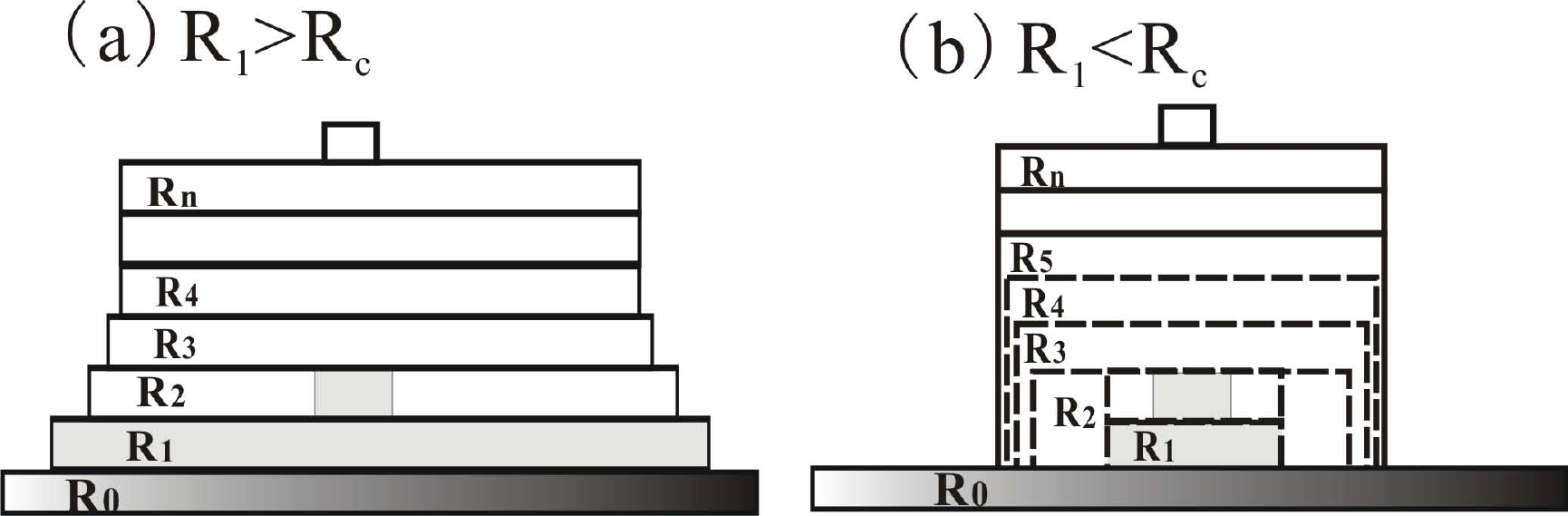}
\caption{Schematic of the two different growth modes from the initial island (grey colors) with radius of $R_1$: (a) $R_1>R_c$; (b) $R_1<R_{c}$. $R_0$ denotes the average radius of the substrate occupied by per island, and $R_n$ is the radius of the topmost layer in the nanorod with $n$ grown atomic layers.}
\label{fig1}
\end{figure}

As discussed in previous section, in heterogenous growth the characteristic radius for the second layer changes   from $R_{c0}$ to $R_{c}$. In homogeneous growth, or in late stage of the heterogenous growth where substrate effect is negligible,  the characteristic radius can vary by changing the deposition rate $F$ or the temperature $T$,  according to Eq.\ (\ref{eq:L2d}). In these cases, $R_{c0}$ and $R_{c}$ are defined as the characteristic radius before and after changing the growth conditions respectively,  $R_0$ and  $R_1$ correspond to the radius of the topmost two layers at the moment that a nucleus forms on $R_1$ after changing the growth conditions.

Two growth scenarios can be identified according to the way that the characteristic radius changes. The first scenario occurs when $R_c$  decreases from $R_{c0}$, $i.e.$, $R_{c}< R_{c0}$. Since $R_{1}>R_{c0}$, it is still larger than $R_{c}$. Thus the third layer forms atop  when ${R_{2}}$ approaches $R_{c}\left( {R_{1}}/{R_{c}}\right) ^{2/7}$. If we assume all the buried layers  cease growing, the radius of each layer is fixed at the moment when it is buried by a new layer. Therefore the radius of the $i$-th layer in a nanostructure with $n$ grown atomic layers, $R_i$, is determined by setting $I(R_{i})=1$,
\begin{equation}
\frac{R_{i}}{R_{c}}= \left( \frac{R_{i-1}}{R_{c}}\right) ^{\frac{2}{7}}
=\left( \frac{R_{1}}{R_{c}}\right) ^{(\frac{2}{7})^{i-1}}, (i=2,3,...,n). \label{eq:Rn2d}
\end{equation}
  Eq.~(\ref{eq:Rn2d}) indicates that $R_i$ decreases rapidly with increasing $i$, until it  approaches $R_{c}$. Correspondingly, the growth morphology changes from a tapered one to a nanorod in several transient layers,  as schematically shown in Fig.~1(a).

Obviously this is merely a limiting case in which some strong screening effects  exist so that the topmost layer dominates the deeper layers in capturing the deposit atoms \cite{steer}. The opposite limiting case is that in which the growth units are equally deposited on the exposed area, thus the buried layers can also grow. In the latter  case island growth leads to the well-known wedding-cake morphology \cite{Krug}.

The realistic situation is that between these two cases, in which only some finite topmost layers are involved in capturing the deposit atoms as a result of a mediate screening effect. To illustrate the screening effect on the island growth, we consider that only  the topmost finite $N_g$ layers keep growing laterally. The quantity $1/N_g$, which is in the range of $0$ to $1$, can be taken as a  measure of  the {\it screening strength}. We have carried out numerical calculations of the rate equations  with different values of $N_g$, the details of which will be  reported elsewhere. We find that when the number of the atomic layers of the island $n$ is smaller than $N_g$, the island grows with the well-known  wedding-cake shape.   When $n$  increases larger than $N_g$, the radii $R_i$ for $i<N_g$ grow gradually to $R_0$,  while for $N_g<i<n-N_g$ $R_i$  approach their stable values after sufficient growth,
 \begin{equation}
\frac{R_{i}}{R_{c}}=X^{\left(\frac{2}{7}\right)^{(i/N_g^2)}}, (N_g<i \leq n-N_g). \label{eq:Ng}
\end{equation}
$X$ is a constant determined by $(R_1/R_c)$ and $N_g$.  For $N_g=1$, $X=(R_1/R_c)^{3.5}$, consistent with Eq.~(\ref{eq:Rn2d}).
The radius $R_i$ decreases with $i$, until it approaches $R_c$ after some transient layers and then remains this value.  The number of the transient layers  is proportional to  $N_g^2$, with coefficient $a$ of the order of magnitude of $10$.  We therefore show that, under a certain screening strength, wedding-cake morphologies ($n<N_g$), tapered morphologies ($N_g<n<aN_g^2$) and nanorods ($n>aN_g^2 $) can be observed successively  during the island growth.

It is thus clear that if the radius of the initial island $R_1>R_c$, the structure finally  approaches a nanorod with a tapered base beneath. The {\it screening strength} only influences the number of the transient layers of convergence, $i.e.$, the atomic layers of the tapered base. Since the nucleation takes place on the topmost monolayer, we refer the nanorod converged from $R_{1}> R_{c}$ as the 2D-ESB-limited nanorod, with radius $R_{2D}$ equals to the characteristic radius of the system, $R_{2D}=R_{c}$.  We term this growth mode as the 2D-ESB-limited one.

The second scenario occurs when   $R_c$ increases from $R_{c0}$ to a value  much larger than $R_{1}$.  In this case the average number of nucleus on top of $A_2$ when $A_2$ covers $A_1$ is $(R_{1}/R_{c})^{5}\simeq 0$. Therefore the topmost two layers can bunch to a bilayer, which then grows laterally from $R_{1}$ to $R_{2}$  till a new nucleus finally forms atop. The process repeats and the growth morphology remains rod-like as shown in Fig.~1(b). The whole nanorod grows laterally as the number of the atomic layers  $n$ increases, fed by the deposited adatom on the top of the nanorod. Therefore,
\begin{equation}
\frac{dA_n}{dt}=\frac{1}{n}FA_n=\frac{1}{n} F\pi R_n^2.
\end{equation}
The average number of nucleus can be obtained by integrating Eq.~(\ref{eq:I0}),
\begin{eqnarray}
I&=& \frac{R^5_{n-1}}{R^5_c} +\int_{A_{n-1}}^{A_{n}} \Omega_{3D}  \frac{dt}{dA} dA \nonumber \\
 &=& \frac{R^5_{n-1}}{R^5_c}+  \frac{7n}{5} \frac{\nu'_{2D}}{\nu'_{3D}}\frac{R_n^5-R^5_{n-1}}{R_c^5}, \label{eq:T3d}
\end{eqnarray}
 where $R_c$ is the same characteristic radius defined in  Eq.~(\ref{eq:L2d}). Note $\Omega_{3D}$ denotes the nucleation rate on $A_n$ after $A_n$ approaches $A_{n-1}$, and  $\nu _{3D}^{\prime }$ represents the interlayer  hopping rate across the multilayer step edges. As discussed above, when $R_{c}$ is much larger  than $R_{n-1}$, the first term in Eq.\ (\ref{eq:T3d}) is nearly zero. The radius of nanorod with $n$ atomic layers $R_n$ can be therefore determined by setting  $I=1$ in Eq.\ (\ref{eq:T3d}),
\begin{eqnarray}
R_n^5 
= R_{n-1}^5 +  \frac{5\alpha }{7n} R_c^5, \label{eq:rn5}
\end{eqnarray}
 where $\alpha =\nu _{3D}^{\prime }/\nu _{2D}^{\prime }=\exp[-(E_{es}^{3D}-E_{es}^{2D})/kT]$ and $0<\alpha <1$.   The radii of the $i-th$  layer $R_i$ can be obtained by iterating  Eq.\ (\ref{eq:rn5}),
\begin{eqnarray}
R_i = R_n= \left[ R_{1}^5 + \frac{5}{7}\alpha R_{c}^5 \sum_{h=2}^{n} \frac{1}{h} \right]^{1/5},(i=2,3,...n) \label{eq:Rn3d}
\end{eqnarray}

The nanorod radius increases until  the difference of $R_{n}$ and $R_{n-1}$ is smaller than one lattice parameter for sufficient large $n$, when it approaches the stable radius of the 3D-ESB-limited nanorod $R_{3D}$. According to Eq. (\ref{eq:rn5}), this happens when
\begin{equation}
n=n_s=\frac{\alpha R_c^5}{7R_{3D}^4}. \label{eq:nst}
\end{equation}
Substituting Eq.~(\ref{eq:nst})  into Eq.\ (\ref{eq:Rn3d}) and replacing the summation with logarithm for large enough $n$, the  stable radius of the 3D-ESB-limited nanorod can be obtained as,
\begin{equation}
R_{3D}=\left[ R_{1}^{5}+\frac{5}{7}\alpha R_{c}^{5}(\gamma-1+\ln \frac{\alpha R_{c}^{5}%
}{7R_{3D}^{4}})\right] ^{1/5},  \label{eq:R3d}
\end{equation}%
wheren $\gamma$ is the Euler's constant.
Since $\alpha R_c^5$ is proportional to $\nu'_{3D}/F$, it is evident that the  nanorod growth in this case is determined by the radius of the initial island $R_1$, the deposition rate and the 3D-ESB.
Larger 3D-ESB ( smaller $\alpha $) facilitates the convergence ($i.e.$ smaller $n_s$) and leads to smaller $R_{3D}$, which is consistent with  a recent Monte-Carlo simulation on copper nanorod growth \cite{Huang08}.

\begin{figure}[t]
\includegraphics[width=8.5cm]{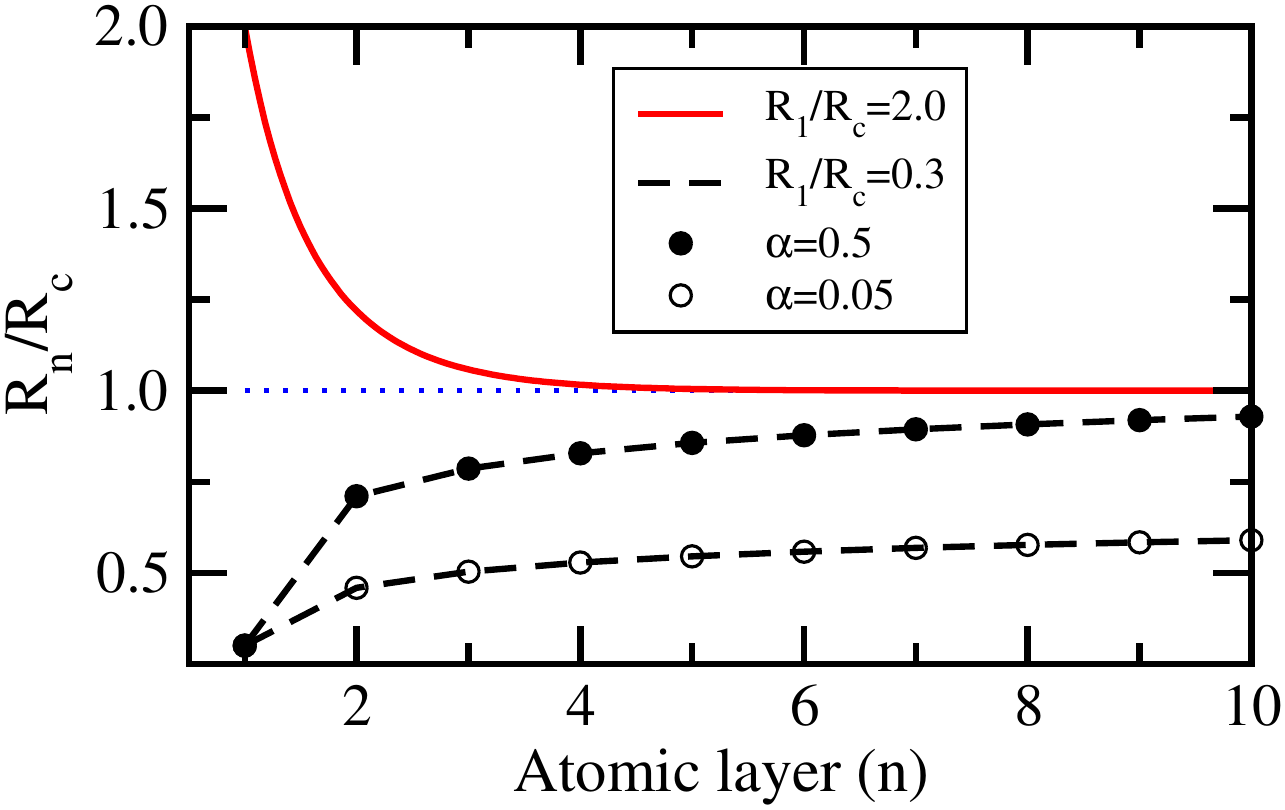}
\caption{The radius of the $n$-th layer $R_n$ in an island with $n$ grown atomic layers, for the cases of $R_1>R_c$ with $N_g=1$ (above the dotted line), and $R_1<R_c$ (below the dotted line), as a function of $n$. The values are calculated according to Eqs.\ (\ref{eq:Rn2d}) and (\ref{eq:Rn3d}). The dotted line guides the characteristic radius $R_{c}$. $\alpha=\nu'_{3D}/\nu'_{2D}$ is the ratio of the hopping rate across the multilayer step to the one across the monolayer step. }
\label{fig2}
\end{figure}

It is worthy to emphasize  that Eq.~(\ref{eq:rn5}) is valid  only  when the first term in Eq.\ (\ref{eq:T3d}) is negligible, and it is physically invalid to extrapolate from Eq.\ (\ref{eq:R3d}) to a $R_{3D}$ larger than $R_c$. Actually, the 2D-ESB-limited growth sets in once $R_{n}$ is increased to $R_{c}$. Therefore the real radius $R_{3D}$ may never exceed $R_{c}$.

For comparison, we show in Fig.~2  the  radius  $R_{n}$ in a nanostructure with $n$ grown atomic layers for the two scenarios, according to  Eqs.\ (\ref{eq:Rn2d}) and (\ref{eq:Rn3d}). It is clear that when the radius of the initial island $R_{1}$ is larger than the characteristic radius $R_{c}$, $R_{n}$ decreases rapidly until it approaches $R_{c}$. Consequently, the growth morphology evolves from a taper-like structure to a nanorod with uniform radius of $R_{c}$. In this scenario, the converged radius is limited by 2D-ESB. When $R_{1}<R_{c}$, $R_n$ corresponds to the nanorod radius. It increases relatively slowly, with a stable radius  smaller than $R_c$, determined  by the 3D-ESB and the radius of the initial island  $R_1$.

\section{V. Experiment Verifications}

\begin{figure}[t]
\includegraphics[width=8.5cm]{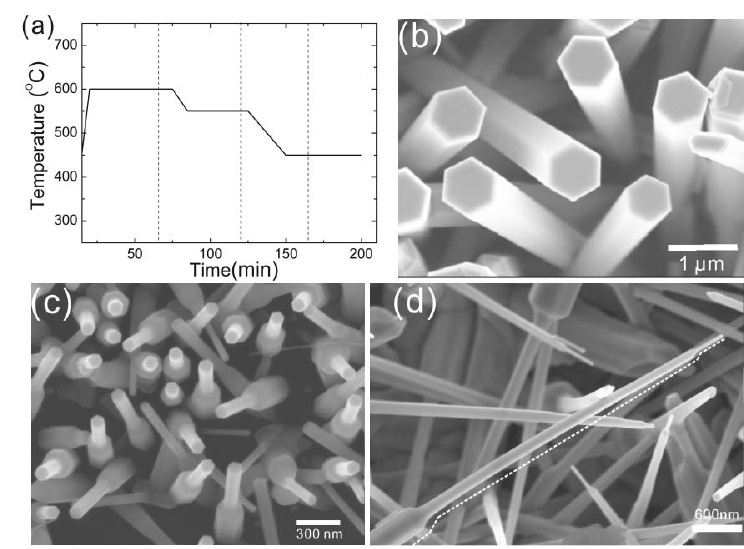}
\caption{ (a) Schematic of the temperature change in experiments, where the three dashed lines correspond to the moments the growth is cut off by large flux of nitrogen. The ZnO products obtained at the three cut off moments are shown  in (b), (c) and (d), respectively. }
\label{fig3}
\end{figure}

Experimentally we take zinc oxide (ZnO) vapor growth as an example to verify the selection of the nanorod radius via varying $R_c$ by changing the growth temperature. For this purpose, unlike conventional ZnO nanorod growth systems where catalysts are usually introduced,  here we establish a physical growth system without using additive chemicals.

The nanorods of ZnO were synthesized catalyst-free in a horizontal tube furnace with programmable temperature control. The pure zinc powder (99.9\% Alfa Aesar) and polished Si(100) substrate were arranged in the same quartz boat and 1.0 cm apart. The growth was carried out with flux of nitrogen controlled as 300 standard cubic centimeter per minute (sccm) and oxygen gas flux as 5 sccm. The temperature in the central section of the furnace was homogeneous, where the quartz boat was placed. In each run of the experiment, the temperature was changed as shown in Fig.~3(a) to control the deposition rate. The growth of nanorods was terminated at different time by sudden increasing the nitrogen flux and cutting off the oxygen supply, as illustrated  by the three dashed lines in Fig.~3(a). In this way, we expect to preserve the growth morphology at the moment that the growth has been terminated.

In our experiments,  the temperature is varied while the flux
of N$_2$ and O$_2$ are kept as  constants for ZnO nanorod growth. The evaporated zinc atoms react with oxygen molecules, and the partial pressure   of the production ZnO is proportional to that of zinc, $ P_\mathrm {ZnO} \propto  P_\mathrm {Zn}/K_p $, where $K_P$ is  the reaction constant. Both $P_\mathrm{Zn}$ and $K_p$  exponentially depend on  temperature,
\begin{eqnarray}
 && P_\mathrm{Zn}\propto \exp(-\frac{B_\mathrm{Zn}}{kT}),~~
  K_P \propto \exp(-\frac{B_K}{kT}),
\end{eqnarray}
 where $B_\mathrm{Zn}= 0.58$ $\mathrm{eV}$ $(6776 \mbox{K})$ \cite{crc} and $B_K=0.21$ $\mathrm{eV}$ $(2474 \mbox{ K})$ \cite{wzl}.
The partial pressure of ZnO can be therefore written as,
\begin{equation}
P_\mathrm{ZnO}= P_0 \exp(-B/kT)
\end{equation}
where $B=B_\mathrm{Zn}-B_K=0.37 \mathrm{eV}$, and $P_0$ is a constant determined by the other growth conditions except of temperature.   The temperature-dependent deposition rate per lattice site can be written as $F=a_{0}^{2}P_{\mathrm{ZnO}}/\sqrt{2\pi mkT}$.

According to Eq.\ (\ref{eq:L2d}), the characteristic radius $R_c$ is
\begin{equation}
R_{c}= sb(kT)^{1/10}\exp (-\Delta E/kT),  \label{eq:zno}
\end{equation}%
where $b=\left[ 7\nu _{0}\sqrt{2\pi m}/(P_{0}a_{0}^{2})\right] ^{1/5}$, $\Delta E=(E_{s}-B)/5$, and $s$ is the geometrical factor associated with the different cross-sectional shapes of nanorod. Equation (\ref{eq:zno}) shows obviously  that $R_c$ can be tuned by changing temperature.  If $\Delta E$ is positive, $R_c$ decreases with decreasing temperature, then the first (2D-ESB limited) growth  scenario is realized, and the radius of nanorod corresponds to the characteristic radius $R_c$. Otherwise if $\Delta E$ is negative, $R_c$ increases with decreasing temperature, and so does the radius of nanorods.

\begin{figure}[t]
\includegraphics[width=8cm]{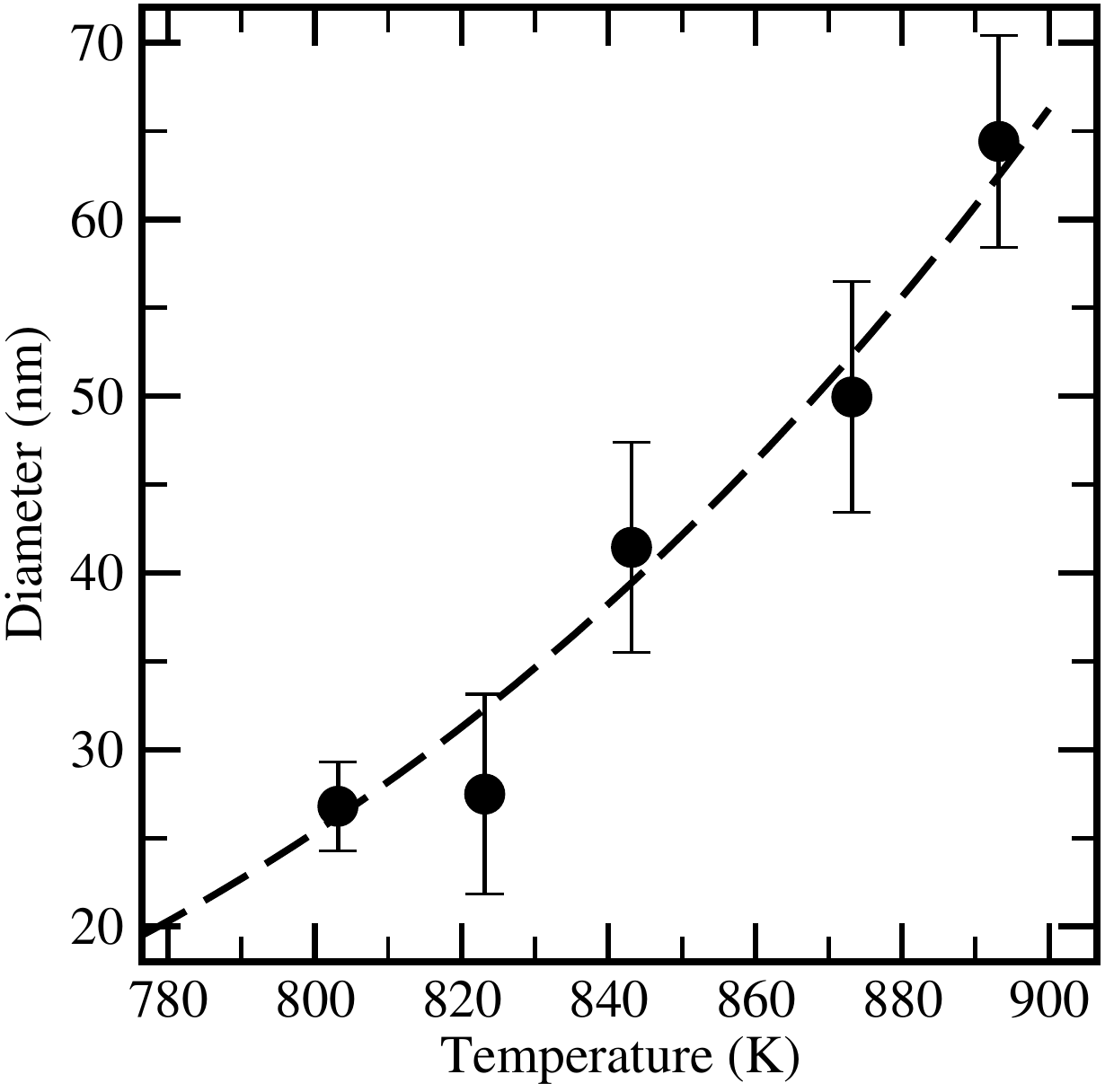}
\caption{ 
The diameter of the second segment nanorod as a function of the corresponding temperature. The dashed curve gives the theoretical fitting results according to Eq.\ (\ref{eq:zno}).}
\label{fig4}
\end{figure}

Now let us  check the variation of morphologies of ZnO nanorods  with step-decreased temperature  as shown in Fig.~3(a).
The morphologies of the nanorods were characterized by a field emission scanning electron microscopy (LEO 1530VP).  Figure 3(b) illustrates ZnO structures grown at constant temperature of $600^{\circ }$C. There is no evident variation of the cross-sectional diameter along the longitudinal direction of the nanorods. When the temperature was decreased from 600 to 550 $^{\circ }$C, a second segment of nanorods appear, the cross-section area of which shrink to  smaller values, as shown in Fig.\ 3(c). The morphologies obtained after two temperature drops are shown in Fig.~3(d), where two evident changes of cross-sectional diameter can be identified on the nanorods as guided by the dashed line.  The cross-section of the ZnO nanorods are of the typical hexagonal one in all cases.

 The experimental observation that the radii of nanorods decrease from one segments to the successive ones  with decreasing temperature suggests that the growth mode is the 2D-ESB-limited one. Therefore the nanorod radius under a certain temperature is expected to be equal to the corresponding $R_c$. In order to eliminate the influence of substrates, the temperature dependence of the radii of the nanorods in the second sections are explored,  while keeping all the other growth conditions as the same.

 In Fig.~4, we plot the circum diameters of the cross-section of the nanorods in the second segments of the structures as shown in Fig.~3(c), as a function of the corresponding temperature.  The dashed curve gives the theoretical fitting results according to Eq.\ (\ref{eq:zno}).   It shows that the theoretical model is in good consistency with the experimental data. The fit value of $\Delta E$ is $0.59$ eV, which leads to $E_{s}$ of about $3.3$ eV. We should point out that this value is only a rough estimate of the effective barrier against the adatom diffusing across a monolayer step edge in zinc oxide system.

\section{VI. Summary}

We have demonstrated that two nanorod growth modes can be realized, depending on a characteristic radius which is determined by the 2D-ESB and the deposition rate. When the radius of the initial island is larger than this characteristic radius, the nanorod radius is 2D-ESB-limited and approaches the characteristic value. Otherwise the nanorod is 3D-ESB limited with a stable radius  smaller than the characteristic radius. We suggest that our results is helpful to select the desired growth modes and control the diameter of nanorods and nanowires.

Although experimental studies on growth of ZnO nanorod with hexagonal cross-section has been reported before, to the best of our knowledge, a quantitative study considering the kinetics in the interfacial growth remains rare. Moreover, the theoretical model proposed here  in fact is a generic one that is not limited to ZnO nanorod growth only. Experimentally, if one can precisely tune the characteristic radius $R_{c}$ for a specific growth system, or choose a desired initial radius by using suitable seeds, either kinds of growth modes can be selected in order to obtain different morphologies and nanorod size. By this means, microscopic informations of 2D-ESB and 3D-ESB can also  be inferred from the experiments.

\section{Acknowledgement}
This work was supported by NSF of China (10974079 and 10874068) and Jiangsu Province (BK2008012), MOST of China (2004CB619005 and 2006CB921804). Z. Zhang acknowledges partial support by USDOE (grant No. DE-FG02-05ER46209, the Division of Materials Sciences and Engineering, Office of Basic Energy Sciences), and USNSF grant No. DMR-0906025.


\end{document}